# A FEW COMMENTS TO A METHOD FOR PRODUCING POSITRONS FOR ILC

A. Mikhailichenko, CLASSE, Ithaca, NY 14853

*Abstract.* We attracted attention to some peculiarities of polarized positron production for Linear Collider (ILC). The conversion system with many targets and low $K$-factor value is our primary interest.

## OVERVIEW

In ILC positron source the undulator scheme is appointed as a baseline. This type of source allows reduction of power dissipation in a target and – the mostly valuable property – it allows generation of *polarized* positrons (and electrons) in quantities $1.5e^+/1e^-$ with average beam polarization ~70%. Helical undulator radiation (UR) generated by particles with energy $E$~150-500 $GeV$ serves as a source of circularly polarized gammas, which converted further into electron-positron pairs in a thin ($0.5X_0$) target. The energy of photons radiated at first harmonic defined by Doppler shift of frequency of wiggling in a Lab frame as

$$\hbar\omega_1 \approx \frac{2\pi\hbar c}{l_u \times (1-\vec{n}\vec{v})} \approx \frac{4\pi\hbar c \gamma^2}{l_u \times (1+K^2+\gamma^2\vartheta^2)} \sim 10\text{-}20 \; MeV,$$

where $\vec{n}$ – is an unit vector towards the observer, $\vec{v}$ – is a particle velocity, $\vartheta$ – is the angle between $\vec{n}$ and $\vec{v}$, $K = eHl_u/2\pi mc^2 = \beta_\perp \gamma$ – where $\beta_\perp$ stands for the transverse velocity of electron normalized by the speed of light, $\gamma = E/mc^2$, $l_u$ –is a period of magnetic field in the undulator; $K$– factor is the mostly important characteristic, which defines the properties of UR. These properties include the harmonic content of radiation, i.e. its spectrum and polarization. So the momentum associated with the transverse oscillation of the particle comes to be $p_\perp = mc\beta_\perp\gamma = mcK$.

Undulator does not require any focusing lenses between sections, at all length the beam is weakly focused by the helical magnetic field of undulator only. The ratio of length with helical field to the total length of undulator is ~0.97.

Beta-function in undulator chosen to be of the order of the undulator length (Code KONN takes this into account).

In undulator the particle's trajectory is a helix with typical parameters for $\gamma\epsilon_x = 10^{-4} cm \times rad$, $\gamma\epsilon_z = 10^{-5} cm \times rad$, $\gamma = 3 \times 10^5$ (150 $GeV$), envelope function $\beta_{x,y}$ ~200 $m$, period of undulator=1 cm, $K$=1, are represented in a Table 1 below.

*Table 1. 150-GeV beam parameters in a helical undulator.*

| Beam size horizontal, $cm$ | $\sqrt{<x^2>} \approx \sqrt{\gamma\epsilon_x \times \beta_x / \gamma}$ | $2.5 \times 10^{-3} cm$ |
|---|---|---|
| Beam size vertical, $cm$ | $\sqrt{<y^2>} \approx \sqrt{\gamma\epsilon_y \times \beta_y / \gamma}$ | $8 \times 10^{-4} cm$ |
| Angular spread in beam, vert. | $\sqrt{<y'^2>} \approx \sqrt{\gamma\epsilon_y / \beta_y / \gamma}$ | $4 \times 10^{-8} rad$ |
| Angular spread in radiation, vert | $\alpha \approx \sqrt{1+K^2}/\gamma$ | $5 \times 10^{-6} rad$ |
| Radius of helix | $r \approx l_u K/\gamma$ | $5 \times 10^{-7} cm$ |

One can see that the radius of helixes, - 0.005 *μm*- is much smaller, than the vertical beam size in undulator, which is~8 *μm*. Angular spread in the beam is ~1% of angular spread of radiation, so the collimation of the photon beam has direct sense here. Installation of focusing lenses in undulator violates domination of angular spread of radiation on angular spread in a beam, so the collimation of undulator radiation for enhancing polarization of radiation becomes meaningless.

## LOW K-FACTOR

Now the idea, that the *K*-factor should be small, attracting attention at last [2], although explanations on why *K*-factor should be small are going since the VLEPP times (1980). For VLEPP, the optimization gave $K \approx 0.3$ (more exactly it was chosen $K^2 = 0.1$).

With reduced *K*-factor the aperture of undulator could be increased. At Cornell, the undulator sections with the length of 3-*m* were developed; The sections of undulator arranged by pairs, so the undulator unit is ~6*m*-long. Sections arranged by pairs for cancellation of the first and second integrals of magnetic fields by relative twist of one section with respect to another one by~$90^o$. That is a peculiarity of helical undulator. Small (~2 *cm*-long) dipole correctors are installed between the pair of sections. In final design we suggested even longer sections.

Vacuumed system runs through all undulator length, pretty much as in SC linacs (or in LHC). Cornell undulator has **8 *mm*** of inner diameter of aperture.

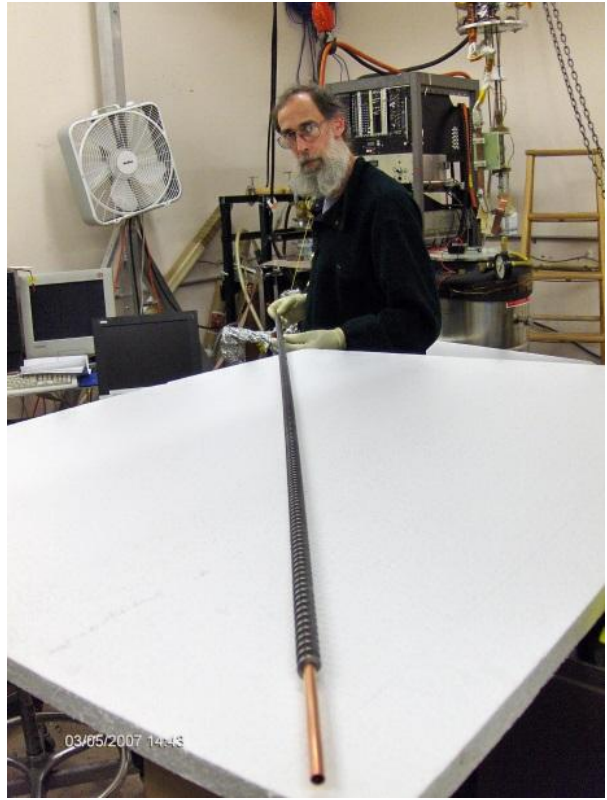

Figure 1. Undulator yoke under test at Cornell. Inner diameter of thin-wall OFC tube is 8 *mm*, period $l_u$ ~1 *cm*.

The energy of gammas-$E_\gamma$ could be made low enough by choosing of appropriate period, so the energy of the first harmonic remains $E_{\gamma 1} < 20$ *MeV*; this means that the absorber (dump) for photons could be designed so that it takes into consideration the fact, that the gamma-neutron production has a threshold (for Al~13 *MeV*, for C~19 *MeV*), so the induced radioactivity could be made low enough. Residual activation could be screened by Borated concrete. Again, low *K*-factor is a preferable option.

The share of first harmonic in total flux is represented in Fig.2.

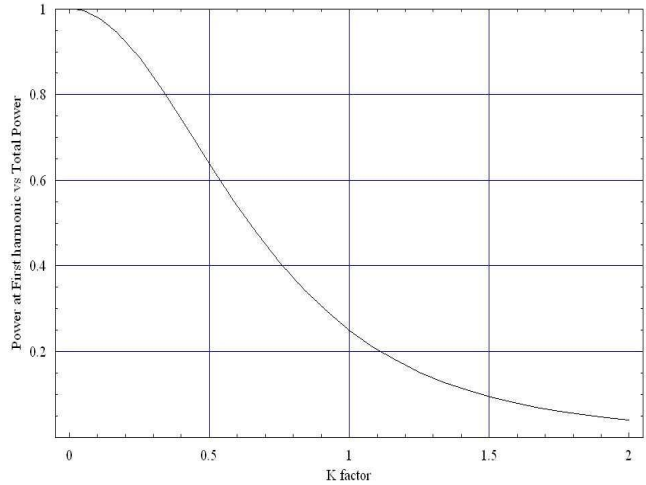

Figure.2. The ratio of power radiated at first harmonic to the total power as a function of *K*-factor.

## POLARIZATION

For intermediate *K*-factor, when the first harmonic is dominating, the function of collimator is in ***reduction*** of second and other harmonics as they have zero intensity at the axis and only 50% polarization in direction of angle $\sim 1/\gamma$ (for second harmonic). By the other words, the photon radiated on second harmonic under angle $\sim 1/\gamma$ has energy equal to the energy of the photon radiated at first harmonic into straight forward direction; but the photon radiated at second harmonic has polarization ~0% only. So if these two photons (radiated on the first harmonic and the one radiated on the second harmonic under angle $1/\gamma$ ) create a positrons with maximal energy, the collection optics cannot resolve (select) the positrons created by these gammas. But now the positron from the photon of second harmonic carries polarization 0% or less.

The procedure of energy selection, **automatically** selects the particles with highest energy **and hence**, with small angular deviation from the axis. So it is desirable to prevent accepting of positrons created by photons of different harmonics. Polarization of created positron as function of its energy $E_+$ (Olsen, Maximon) has the following form:

$$\vec{V} = \chi_2 \times [f(E_+/E_g) \times \vec{n}_\parallel + g(E_+/E_g) \times \vec{n}_\wedge] = \vec{V}_\parallel + \vec{V}_\wedge, \qquad (1)$$

where $\chi_2 \approx (1 - g^4 J^4)/(1 + g^4 J^4)$ - a circular polarization of initial gamma-beam-is a function of collimation diameter linked by the angle of radiation $\vartheta$; dependence of $\chi_2$ on angle for the first and second harmonics is the same. Functions *f* and *g* are represented in Fig.3. If the angle of radiation is increased, the energy of quanta drops accordingly, so the maximal energy of created positron. So the separation of particle by energy is a primary factor for enhancement of polarization.

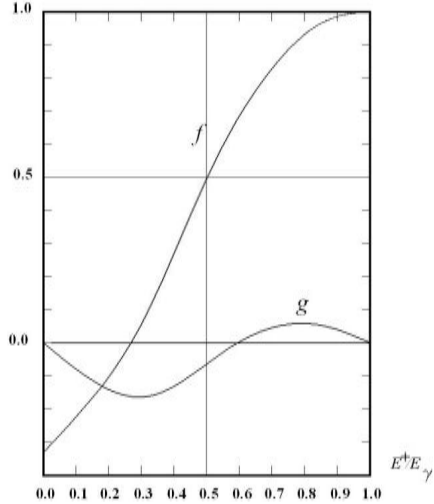

Figure 3. Functions *f* and *g*, (Olsen Maximon).

One comment of usage of back-scattered laser radiation for creation of high-energy photons is the following. As the radiation of electron in a back- scattered radiation from a laser, can be described in a same way as the radiation from an undulator, while the energy of secondary photon is much lower than the energy of electron, the recommendation for lowering *K* factor automatically fulfilled here due to practical limitation of power achievable in a laser system.

Operation with low *K*-factor in E-166 experiment with *K*~0.17 (SLAC, 2005) together with selection of energy of the secondary positrons by two-magnet spectrometer delivered the polarization **measured** ~85-95% .

So the polarization appears not as result just collimation of undulator radiation (pretty typical statement). As it could be seen from (1) collimation just enhances the first factor $x_2$, but the functions *f* and *g* play more important role. The full picture is the following. The higher harmonics have less than 50% polarization at the photon energy around the first harmonic (or even negative values). The energy of quanta radiated at harmonic is a function of angle between the axis of undulator and the line towards the direction of radiation. But all harmonics besides the first one have zero intensity on the axis of undulator (forward direction, $\vartheta=0$). So the collimator rejects the higher harmonics (the second one mostly). As it could be seen from Fig.2, the first harmonic reaches its maximum at *K*~0.7, where it's content in a total intensity reaches ~50%. So there is no any sense to have *K*-factor bigger, than 0.7 in any case, but it should be lower. At VLEPP with *K*-factor ~0.3 the collimation was not required at all, as the capturing system with Lithium lens had narrow energy acceptance for collected positrons. So collecting positrons around maximal possible value of energy (by proper tuning) automatically selects high positrons with high level of polarization. The collimation of gammas helps in reduction of power, dissipated in a target however.

## POSITRON YIELD

In Table 2 the parameters of conversion system at different energy of driving beam (electron or positron) is represented for few different energies of the primary electron (or positron) beam. Calculations done with KONN; focusing provided by the Lithium lens.

*Table 2*. Efficiency and polarization achievable with undulator scheme (KONN).

| Beam energy, GeV | 150 | 250 | 350 | 500 |
|---|---|---|---|---|
| Length of undulator, m | 180 | 200 | 200 | 200 |
| **K factor** | **0.45** | **0.44** | **0.35** | **0.27** |
| Period of undulator, cm | 1.0 | 1.0 | 1.0 | 1.0 |
| Distance to the target, m | 150 | 150 | 150 | 150 |
| Radius of collimator, cm | 0.049 | 0.03 | 0.02 | 0.02 |
| Emittance, cm·rad | $10^{-9}$ | $10^{-9}$ | $10^{-9}$ | $10^{-9}$ |
| Bunch length, cm | 0.05 | 0.05 | 0.05 | 0.05 |
| Beta-function, m | 400 | 400 | 400 | 400 |
| Thickness of the target/$X_0$ | 0.57 | 0.6 | 0.65 | 0.65 |
| Distance to the lens, cm | 0.5 | 0.5 | 0.5 | 0.5 |
| Radius of the lens, cm | 0.7 | 0.7 | 0.7 | 0.7 |
| Length of the lens, cm | 0.5 | 0.5 | 0.5 | 0.5 |
| Gradient in lens, MG/cm | 0.065 | 0.065 | 0.08 | 0.1 |
| Wavelength of RF, cm | 23.06 | 23.06 | 23.06 | 23.06 |
| Phase shift of crest, rad | -0.29 | -0.29 | -0.29 | -0.29 |
| /Distance to RF str., cm | 2.0 | 2.0 | 2.0 | 2.0 |
| Radius of collimator[†], cm | 2.0 | 2.0 | 2.0 | 2.0 |
| Length of RF str., cm | 500 | 500 | 500 | 500 |
| Gradient, MeV/cm | 0.1 | 0.1 | 0.1 | 0.1 |
| Longitudinal field, MG | 0.045 | 0.045 | 0.045 | 0.045 |
| Inner rad. of irises, cm | 3.0 | 3.0 | 3.0 | 3.0 |
| Acceptance, MeV·cm | 5.0 | 5.0 | 5.0 | 5.0 |
| Energy filter, E > -MeV | 54 | 74 | 92 | 126 |
| Energy filter, E < -MeV | 110 | 222 | 222 | 250 |
| Efficiency, $e^+/e^-$ | 1.5 | 1.5 | 1.5 | 1.5 |
| **Polarization, %** | **69** | **78** | **78** | **73** |

† Collimator at the entrance of RF structure

One can see than *K*<0.45 could satisfy all needs. **Some collimation required for reduction of higher harmonics content in energy spectrum around the first harmonic.**

## FEW TARGETS [1]

- Mostly of ~100 *kW* (for maximal intensity of ILC collider) of energy associated with the gamma-beam is deposited *in the gamma-beam dump*. Only ~16% of this 100 *kW* power becomes deposited in a target. So *few targets* could be installed in series, and positron collection could be carried from each target and combined further in a longitudinal phase-space [1]. The length of the undulator could be reduced accordingly. (Probably, practical number of targets could be up to 3-4).

Combining on positron bunches is going in longitudinal phase-space. The difference of path length distances could be adjusted so the positron bunches fall into neighboring RF buckets of main radiofrequency system of damping ring, see Fig.4. This scheme is especially useful if the conversion system installed at the end of Linac.

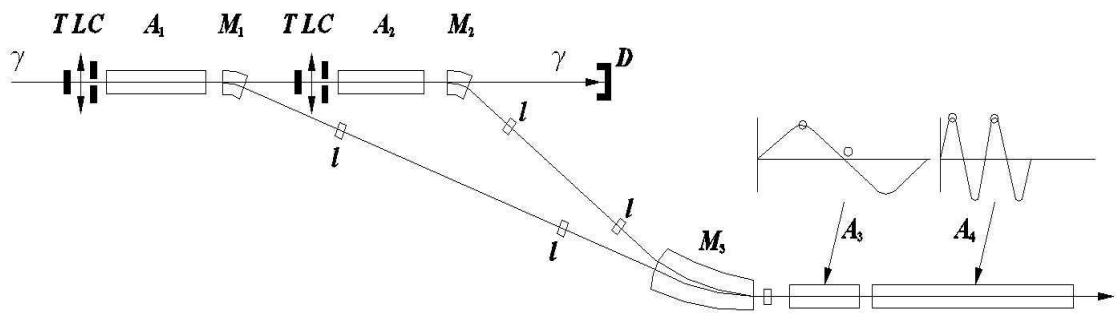

Figuire 4. Example of two-target system [1]. *T*—stands for Target, *L*—stands for Lens, *C*—is a Collimator, *A*—stands for accelerator structure, *M*—for a magnet, *l*—stands for a focusing lens.

The energy provided by acceleration structures *A*1 and *A*2 are slightly different, *A*1>*A*2. Difference $A_1$-$A_2$ should be big enough to separate the beams with reasonable energy spread. In acceleration structure A3, bunches are going in different phases. For given $E\gamma$ the energy spread is less than ½ $E\gamma$~10 MeV, so the difference $A_1$-$A_2$ of 30 *MeV* is big enough.

Basically, the energy selection arranged by screening the positrons at place(s) where the dispersion function reaches its local maximum.

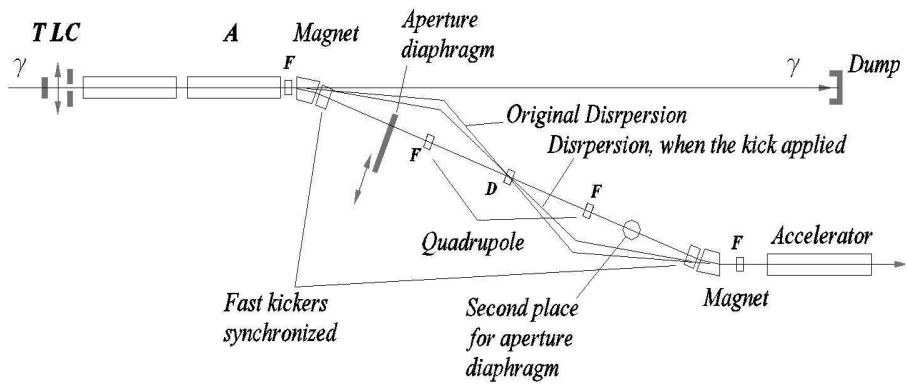

Figure 5. Regulated energy-selection system. This system could be used if the positrons are used in undulator as a primary beam instead the electrons one.

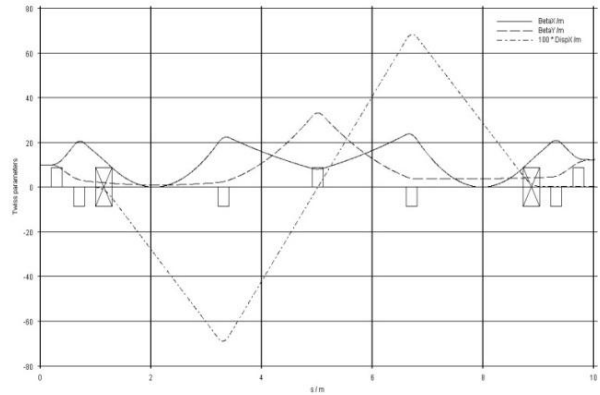

Figure 6. Envelope functions for the optics from Figs.3-4.

The old-fashion tetrode amplifier is appointed here; otherwise a solid-state amplifier of equivalent pulsed power could serve here as well.

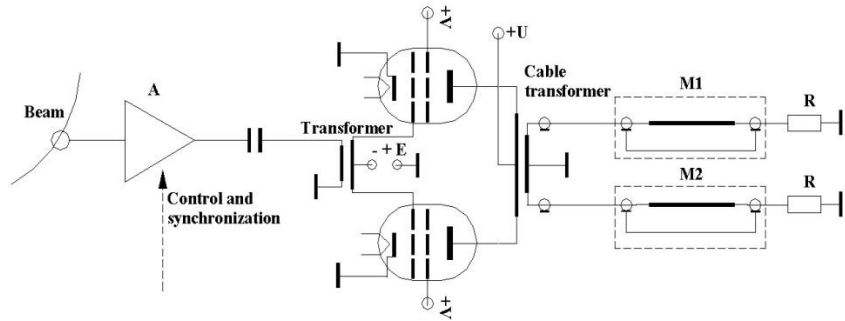

Figure 7. Tetrode amplifier for the fast feedback.

Fast feedback required if the positrons generated by positrons (i.e. the undulator located in the positron wing); otherwise it is not required.
• The active length of undulator for single target required ~20-180 *m* for the energy 120-500*GeV*. For 20*m*-long undulator $K$~1 required, but polarization here <30%.
• Undulator radiation could be created by positrons as well as by electrons, so this might help in optimization of allocation of elements of beam tract; in this case undulator located in the positron line.

## MANIPULATION BY THE BUNCH LENGTH

Do not forget to look at the bunch length, as it might be too short for the damping ring - the instabilities could be excited there. The length of positron bunch right after the target is about the same as the length of the high-energy primary bunch running to the IP. So on the way of positron bunch to the damping ring, some decompression of length is required more likely.
In contrast, on the way to the main linac the bunch length compression required. Such system considered a long time ago [3]. Bunch compression on a basis of wiggler *is extremely ineffective*. Really, the lengthening as a function of bending angle is

$$\Delta l \approx \frac{\rho \varphi^3}{3} \frac{\Delta p}{p},$$

where $\rho$ is a bending radius, $\varphi$ is a bending angle. For the wiggler, $\varphi \sim 2K/\gamma$ x (Number of periods) which remains much smaller, than the bend in a magnet with $\varphi \sim 1$. Radiation in a wiggler is a negative factor also. Probably there is confusion between changing the length of trajectory in the wiggler field (straight line to the sin-like trajectory) and the difference in the path lengths between equilibrium trajectory (also sin-like) and the trajectory for the particle with slightly different momenta.

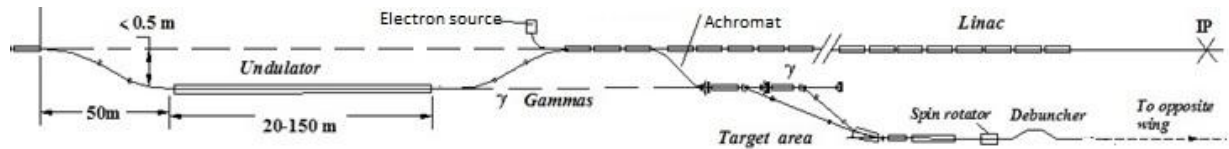

Figure 8. Schematics of conversion system.

Implementation of multi-target system in a collider pattern requires just a little extra space, but undulator could be two times (at least) shorter.

## COMMENT ON SPIN-FLIPPING SYSTEM

For the spin flipping system with two solenoids having oppositely directed magnetic field (which is not the best idea, however), the offset of axes of solenoids could be as small as ~5 *cm* (or even smaller), so these solenoids **can share the same cryostat having combined cold mass**. System equipped with bellows, so for the regimes with some definite polarization, which does not require fast spin-flip, the system temporary could be shifted mechanically, pretty much in the same way as it was done in E-166 experiment at SLAC (2005).

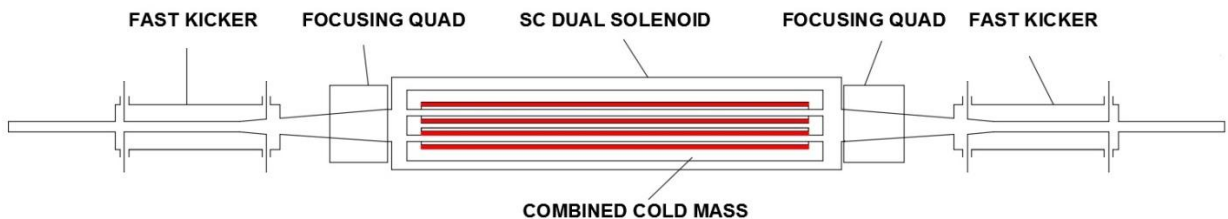

Figure 9. Arrangements for fast switching of polarization. Additional vertical focusing lenses are not shown;

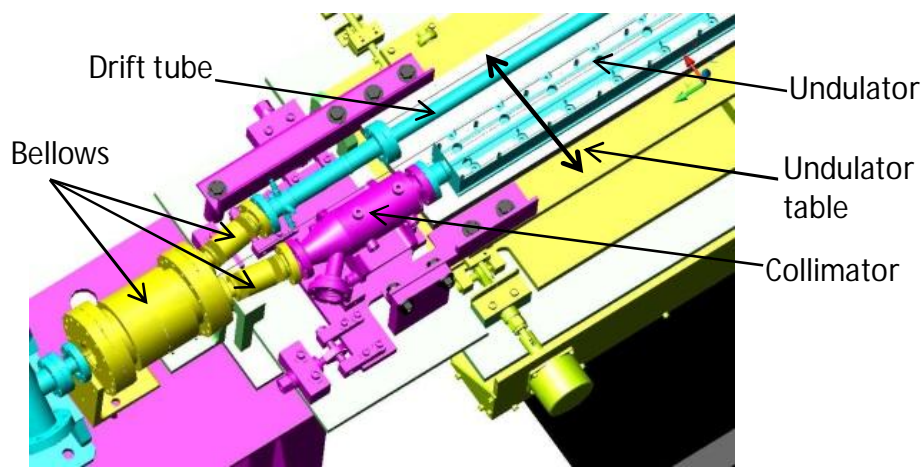

Figure 10. Arrangement of switching between beam passage through the undulator and through the drift tube.

This system from Fig.10 was used in E-166 experiment at SLAC. The same three-bellow system installed at the other side of undulator table. Fat double-edge arrow shows the direction of motion of undulator table.

### EVACUATION OF HEAT FROM SPINNING TARGET DISC BY LIQUID METAL

The idea of contact heat evacuation (W.Gay *et.al.*) could be moved further by suggestion to use the **liquid metal jet** (In/Ga alloy). In addition, this jet can spin the target wheel as a kind of turbine. Simple feedback stabilizes the rotation speed.

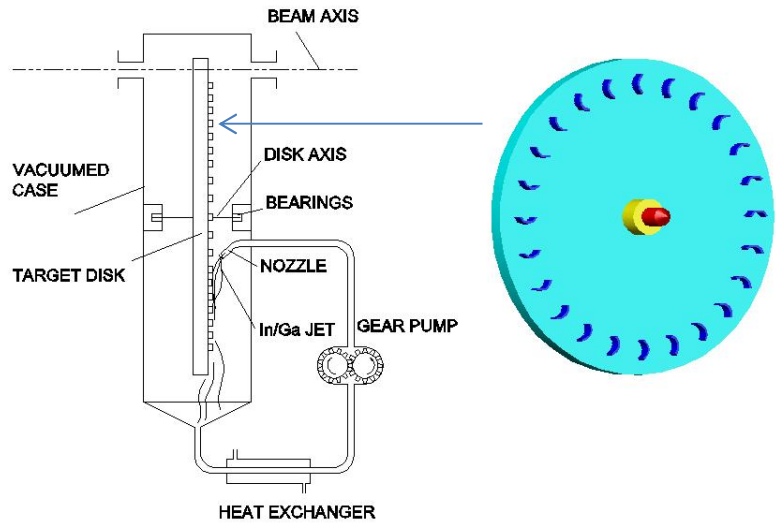

Figure 11. Liquid metal jet spins the target wheel and evacuates the heat deposited in target.

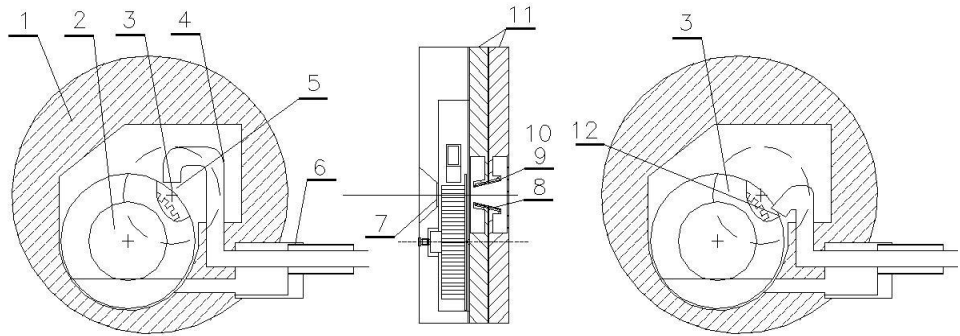

Figure 12. Liquid metal target for VLEPP. Variant 1. 1-Titanium case, 2-is the teethed wheel, 3-is the target focusing point, 4-is the nozzle, 5-is the Mercury jet, 6-is the feeding tubes, 7-is secure Titanium foil, 8-is the conically shaped lens, 9-is the volume with liquid Lithium, 10-is Beryllium made flange, 11-are the current leads made from Titanium. Variant 2. 3 is the target focusing point, 12-is the nozzle. Diameter in Lithium cone at the exit of lens is ~1 *cm*.

This idea was reserved for use in VLEPP (1986) [1]; In first variant the liquid jet just spinning the target disk and evacuating the heat; in the second variant –the jet itself serves as a target and spins thin protecting disk.

# FULL POWER COLLIMATOR

The undulator (as well as the accelerating structure) should be protected against direct hit of primary beam train. This device could be used as a beam stop also. Total energy carrying by the beam train of 2625 bunches populated with $2 \times 10^{10}$ electrons (positrons) with energy 350 *GeV* is

$$Q \approx 350 \times 10^9 \times 1.6 \times 10^{-19} \times 2 \times 10^{10} \times 2625 \approx 2.94 [MJ].$$

If all this energy will be deposited in a 1 kilogram of Gallium, then the temperature rise will be

$$\Delta T = Q/mc_p \approx 2.4 \times MJ/1/0.34[kJ/kg/^oC \approx 7000^oC.$$

This means that some fraction of Gallium will be transformed into vapor at boiling temperature ~2200$^o$C. For vaporization process the latent heat of vaporization should be taken into account, which is *q*~3.6 *MJ/kg*. It looks that only small fraction of liquid metal will be transformed into vapor.

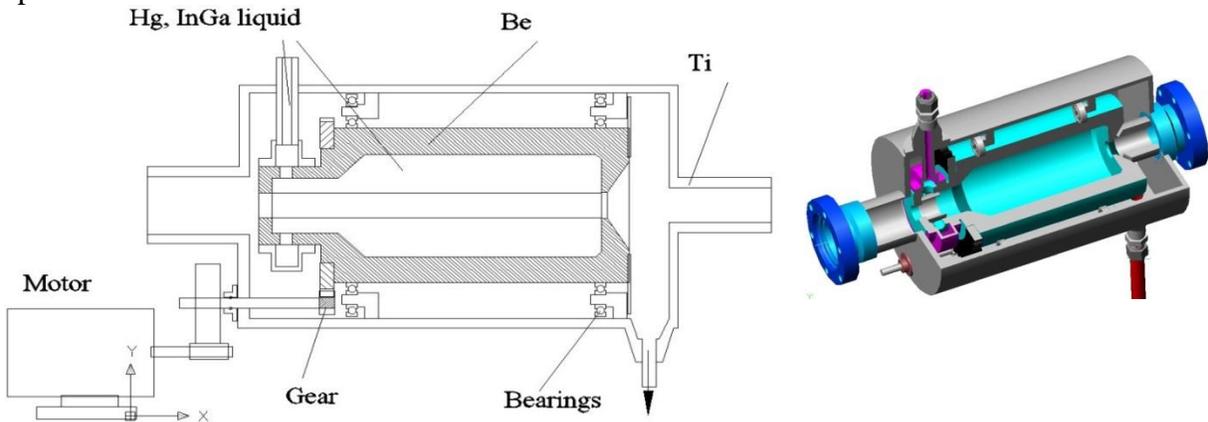

Figure 13. The full power collimator. The length along the beamline is not in scale. This type of collimator could be installed in front of undulator and along the Linac itself.

So as the main problem with collimator is destruction of its walls hit by the train or even by a single bunch, the inner surface of collimator should be an **open surface of liquid metal**. So this liquid metal surface could be formed as a result of *centrifugal force*. Motor is spinning the central bottle in Fig.13 with angular speed ~3000 turns/min.

# ONE COMMENT ON THE HYBRID TARGET CONVERSION SYSTEM

The schematic here is the following. In the first target, the primary electron beam generates the photon(s) by Bremsstrahlung. Electrons are separated away before the second target by the magnet. The photons created in the first target illuminating the second (thin) target, where the positron-electron pairs are created.

The argument here is the following. For the thickness of first target ~$X_0$, each electron generates there ~1 photon in all spectra. Meanwhile in undulator each electron generates ~100 photons in *narrow spectra*. So, to be compatible, the system with hybrid target should compensate this factor ~100-300.

Usage of channeling effects gives hope for compensation factor of 5 maximum (looks too optimistic, however), so the residual factor 20-60 should be compensated by *repetition rate*. So 300 *Hz* system looks more or less adequate here. The polarization of secondary positron beam is absent however.

**EXTREMELY EFFECTIVE POSITRON CONVERSION SYSTEM AT CORNELL**

Let us consider the conversion device used at Cornell accelerator known as mostly efficient in its class as example. Focusing of positrons is arranged by a bi-layer coil with partial flux concentrator inside this coil, Fig.14.

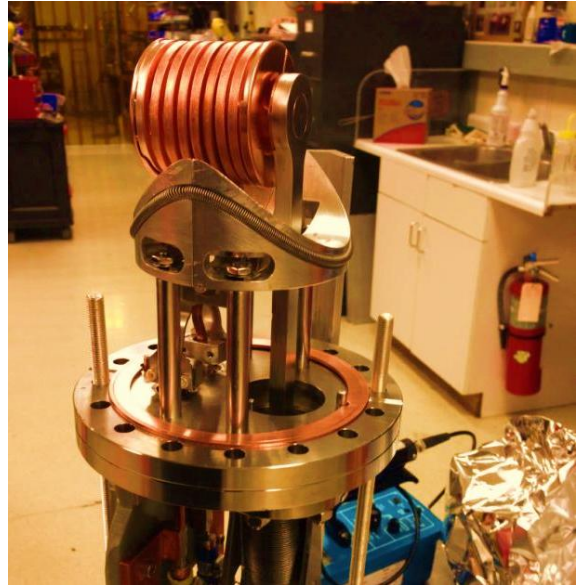

Figure 14. Special collar with spring contact closes the input in a cylindrical case.

W target mounted on the pendulum for the possibility to be moved out of the beam pass. The target is water-cooled. Main effort was applied to design a system with absence of transverse kick. Conversion system delivers >160 *mA/min* e+ in CESR (after acceleration in a buster synchrotron, ejection, injection in CESR); it operates at 50 *Hz*. As the average radius of CESR is~70 *m*, this corresponds to $\sim 3.7 \times 10^{12} e^+/\sec$, which is 1% of the rate, required by ILC $(3.9 \times 10^{14} e^+/\sec)$.

## REFERENCES

[1] A.Mikhailichenko, PhD Thesis, 1986. Translation coud be found in
   http://www.cornell.edu/public/CBN/CBN02-13/DISSERT.pdf,
[2] Bulyak, Shulga., this Workshop.
[3] V.Anashin et.al., "*The Prototype of Damping Ring and the Buncher for the VLEPP Project*", XIII Int. Conf. on High Energy Acc., Novosibirsk, 1986, Proc., pp. 159-163.